\documentclass[twocolumn,showpacs,amsmath,amssymb,prl,superscriptaddress]{revtex4-1}
\usepackage{graphicx}
\usepackage{dcolumn}
\usepackage{bm}

\begin{document}

\title{Spontaneous fission
lifetimes from the minimization of self-consistent collective action}

\author{Jhilam Sadhukhan}
\email{jhilam@utk.edu}
\thanks{on leave of absence from VECC, Kolkata, India}
\affiliation{Department of Physics and Astronomy, University of Tennessee, Knoxville, Tennessee 37996, USA}
\affiliation{Physics Division, Oak Ridge National Laboratory, P. O. Box 2008, Oak Ridge, Tennessee 37831, USA}
\author{K. Mazurek}
\affiliation{Department of Physics and Astronomy, University of Tennessee, Knoxville, Tennessee 37996, USA}
\affiliation{Physics Division, Oak Ridge National Laboratory, P. O. Box 2008, Oak Ridge, Tennessee 37831, USA}
\affiliation{The Niewodniczanski Institute of Nuclear Physics - PAN, PL-31-342 Krakow, Poland}
\author{A. Baran}
\affiliation{Department of Physics and Astronomy, University of Tennessee, Knoxville, Tennessee 37996, USA}
\affiliation{Physics Division, Oak Ridge National Laboratory, P. O. Box 2008, Oak Ridge, Tennessee 37831, USA}
\affiliation{Institute of Physics, University of M. Curie-Sk{\l}odowska, ul. Radziszewskiego 10, 20-031 Lublin, Poland}
\author{J. Dobaczewski}
\affiliation{Department of Physics and Astronomy, University of Tennessee, Knoxville, Tennessee 37996, USA}
\affiliation{Physics Division, Oak Ridge National Laboratory, P. O. Box 2008, Oak Ridge, Tennessee 37831, USA}
\affiliation{Institute of Theoretical Physics, Faculty of Physics, University of Warsaw, ul. Ho$\dot{z}$a 69, PL-00-681 Warsaw, Poland}
\affiliation{Department of Physics, P.O. Box 35 (YFL), University of Jyv\"askyl\"a, FI-40014  Jyv\"askyl\"a, Finland}
\author{W. Nazarewicz}
\affiliation{Department of Physics and Astronomy, University of Tennessee, Knoxville, Tennessee 37996, USA}
\affiliation{Physics Division, Oak Ridge National Laboratory, P. O. Box 2008, Oak Ridge, Tennessee 37831, USA}
\affiliation{Institute of Theoretical Physics, Faculty of Physics, University of Warsaw, ul. Ho$\dot{z}$a 69, PL-00-681 Warsaw, Poland}
\author{J. A. Sheikh}
\affiliation{Department of Physics and Astronomy, University of Tennessee, Knoxville, Tennessee 37996, USA}
\affiliation{Physics Division, Oak Ridge National Laboratory, P. O. Box 2008, Oak Ridge, Tennessee 37831, USA}

\date{\today}

\begin{abstract}
The spontaneous fission lifetime of $^{264}$Fm  has been studied within nuclear density functional theory
by minimizing the collective action integral for fission in a two-dimensional quadrupole collective space representing
elongation and triaxiality. The collective potential and inertia tensor are obtained  self-consistently
using the Skyrme energy density functional and density-dependent  pairing interaction.
The resulting spontaneous
fission lifetimes are compared with the static result obtained with the minimum-energy pathway. We show that fission pathways strongly depend  on assumptions underlying collective inertia. With the non-perturbative  mass parameters, the dynamic fission pathway becomes strongly triaxial and it approaches the static fission valley. On the other hand,
when the standard perturbative cranking inertia tensor is used, axial symmetry is restored along the path to fission; an effect that is an artifact of
the approximation used.
\end{abstract}

\pacs{24.75.+i, 25.85.Ca, 21.60.Jz, 21.60.Ev, 27.90.+b}


\maketitle


\textit{Introduction.}---The spontaneous fission (SF) of a nucleus plays important role in
many areas of science and applications \cite{Wagemans,KrappePom,ShultisFaw}. In particular, it determines
the stability of the heaviest and superheavy elements \cite{Oganessian07,Sta13} and it  impacts the formation of heavy elements at the final stages of the r-process through the recycling mechanism \cite{Arn07,Pan10,ErlLan12}.
Therefore, a capability of theory to predict SF lifetimes in a reliable way  is essential.

The main ingredients for a theoretical determination of SF lifetimes are the collective potential and inertia
tensor. For heavy systems, these quantities can be calculated by using the self-consistent mean field theory based on the energy density functional \cite{Ben03}. The potential energy surface (PES) is obtained by solving constrained
Hartree-Fock-Bogoliubov equations (HFB) in a multidimensional space of collective coordinates. The collective inertia (or mass) tensor  is obtained from the self-consistent densities by employing the adiabatic time-dependent HFB  approximation (ATDHFB) \cite{Baranger1978123,Doba81,Bar11}. Since SF is a quantum-mechanical tunneling process and the fission barriers are usually both high and wide, the SF lifetime is obtained  semi-classically by minimizing the fission action integral in the collective space.

The main objective of this work is to study SF  by combining the microscopic
density functional input with the sophisticated action minimization
techniques. We demonstrate that the predicted SF pathway strongly depends on the choice of the collective inertia. In particular, in the commonly used perturbative cranking approximation, the variations of mass parameters due to level crossings (configuration changes) are underestimated; this results in an artificial  restoration of axial symmetry in the region of the first barrier that is broken in a static and non-perturbative approaches.

\textit{Model.}---In a semi-classical approximation, the SF half-life  is given by~\cite{Baran198183,Baran19788}
$T_{1/2}=\ln2/(nP)$, where $n$ is the number of assaults on the fission barrier per unit time and $P$ is the penetration probability given by
\begin{equation}\label{penertration}
P=\left(1+\exp{[2S(L)]}\right)^{-1}.
\end{equation}
In Eq.~(\ref{penertration}) $S(L)$ is the fission action integral calculated along the one-dimensional fission path
$L(s)$ pre-selected in the multidimensional collective space:
\begin{equation}
\label{action-integral}
S(L)=\int_{s_{\rm in}}^{s_{\rm out}}\frac{1}{\hbar}\sqrt{2\mathcal{M}_{\text{eff}}(s)
\left(V_{\text{eff}}(s)-E_0\right)}\,ds,
\end{equation}
where $V_{\text{eff}}(s)$ and
$\mathcal{M}_{\text{eff}}(s)$ are the effective potential energy and
inertia along the fission path $L(s)$, respectively. $V_{\text{eff}}$ can be obtained by
subtracting the vibrational zero-point energy $E_{\text{ZPE}}$ from the total Hartree-Fock-Bogoliubov energy $E_{\text{tot}}$. Integration limits $s_{\rm in}$
and $s_{\rm out}$ are the classical inner and outer turning points, respectively,  defined by $V_{\text{eff}}(s)=E_0$  on the two extremes of the fission path. The collective ground state energy is $E_0$, and $ds$ is the element
of length along $L(s)$.  A one-dimensional path $L(s)$ can be defined in the multidimensional collective space by
specifying the collective variables $q_i(s)$ as functions of path's length $s$. The most probable fission path corresponds to the minimum of  $S(L)$ \cite{Bra72,Ska08}. The expression for $\mathcal{M}_{\text{eff}}$ is~\cite{Baran198183,Baran19788,Baran05}:
\begin{equation}
\label{eff-mass}
\mathcal{M}_{\text{eff}}(s)=\sum_{ij}\mathcal{M}_{ij}\frac{dq_{i}}{ds}\frac{dq_{j}}{ds},
\end{equation}
where $\mathcal{M}_{ij}$ are the components of multidimensional collective inertia tensor.

In this pilot study, we consider the SF of $^{264}$Fm. This nucleus is predicted to undergo a symmetric fission due to shell effects in the doubly magic nucleus $^{132}$Sn \cite{Sta09}. Consequently, we
consider  a two-dimensional
collective space of mass (isoscalar) quadrupole moments $Q_{20}\equiv Q_1$
(elongation) and $Q_{22}\equiv Q_2$ (triaxiality) defined as in Table 5 of
Ref.~\cite{doba04}. To compute the total energy $E_{\text{tot}}$
and inertia tensor $\mathcal{M}_{ij}$, we employed the
symmetry-unrestricted  HFB solver HFODD
(v2.49t)~\cite{Schunck2012166}.

Throughout this work, we closely follow  Refs.~\cite{Bar11,Sta13}. Namely,
in the particle-hole channel  we use the Skyrme energy density functional SkM$^*$ \cite{Bartel198279}. The particle-particle interaction  was approximated by the density-dependent mixed pairing force \cite{doba02}.

The zero-point energy was estimated by using the Gaussian overlap
approximation~\cite{Staszczak1989589,Bar07,Sta13}.  The collective ground state energy $E_0$ was arbitrarily assumed to be equal to 1\,MeV and, consistently, the value of $n$ comes out to be $10^{20.38}s^{-1}$~\cite{Baran05}.
As discussed in Ref.~\cite{Ska08}, the  action integral
(\ref{action-integral}) can be derived in the adiabatic limit of the imaginary-time-dependent  HFB theory. While the full-fledged calculation of ATDHFB inertia still needs to be carried out, in this work
we use  the non-perturbative cranking approximation~\cite{Bar11}:
\begin{equation}
\label{cranking-mass}
\mathcal{M}^{C}_{ij}=\frac{\hbar^2}{2\dot{q}_i\dot{q}_j}\sum_{\alpha\beta}\frac{\left(F^{i*}_{\alpha\beta}F^{j}_{\alpha\beta}+
F^{i}_{\alpha\beta}F^{j*}_{\alpha\beta}\right)}{E_{\alpha}+E_{\beta}},
\end{equation}
where  $q_i = \langle Q_i\rangle$, $\dot{q}_i$ represents the time derivative of
$q_i$s, the double sum runs
over all one-quasiparticle states $|\alpha\rangle$ and $|\beta\rangle$ which also include the Kramers' degenerate states, and $E_{\alpha}$ are
one-quasiparticle energies. The matrices $F^i$ are given by~\cite{Bar11}:
\begin{equation}
\label{equation-F}
\frac{F^{i*}}{\dot{q}_i}=
 A^T\frac{\partial\kappa^*}{\partial q_i}A
             +A^T\frac{\partial\rho^*}{\partial q_i}B
-B^T\frac{\partial\rho}{\partial q_i}A
             - B^T\frac{\partial\kappa}{\partial q_i}B,
\end{equation}
where $A$ and $B$ are the standard Bogoliubov matrices,
and $\rho$ and
$\kappa$ are particle and pairing density matrices, respectively, determined
in terms of $A$ and $B$. Derivatives of the density matrices with respect
to deformations were calculated by employing the three-point Lagrange
formula~\cite{Gia80,Yuldashbaeva19991}. Finally, the total $\mathcal{M}^{C}$
is obtained by adding the neutron and proton contributions.

For comparison,
we also study the perturbative-cranking inertia tensor~\cite{Bar11,Staszczak1989589}, for which the
derivatives of density matrices are determined
perturbatively in terms of matrix
elements of quadrupole moments. This simplification leads to the
expression~\footnote{Here we corrected misprinted
equations (56) and (60) of Ref.~\cite{Bar11}.}:
\begin{equation}
\label{pc-mass}
\mathcal{M}^{C^{\rm p}}=\hbar^2[M^{(1)}]^{-1}[M^{(3)}][M^{(1)}]^{-1},
\end{equation}
where the energy-weighted moment tensors $M^{(k)}$ are defined as
\begin{equation}
M^{(k)}_{ij}=\sum_{\alpha\beta}
\frac{\langle 0|{Q}_i|\alpha\beta\rangle\langle\alpha\beta|{Q}_j^{\dagger}|0\rangle}
     {\left(E_{\alpha}+E_{\beta}\right)^k},
\label{moment-tensors}
\end{equation}
with $|\alpha\beta\rangle$
being two-quasiparticle wave functions. Similar to the non-perturbative inertia, the total moments $M^{(k)}$s were evaluated by adding contributions from neutrons and protons. The perturbative-cranking collective masses $\mathcal{M}^{C^{\rm p}}$ have been widely used  \cite{Sta13,Baldo13,Giuliani13,Warda02,Baran05,Smo93,Smo95,Ghe99,War12,Del06} to calculate SF lifetimes.

It is important to remark that  rapid variations in $\mathcal{M}^{C}_{ij}$ are expected in the regions of configuration changes (level crossings) due to
strong variations of density derivatives in (\ref{equation-F}) associated with structural rearrangements. Such variations are quenched in perturbative moment tensors (\ref{moment-tensors}) as the matrix elements
$\langle 0|{Q}_i|\alpha\beta\rangle$ are actually {\it reduced} in  level-crossing regions.

\begin{figure}[htb]
\includegraphics[width=\columnwidth]{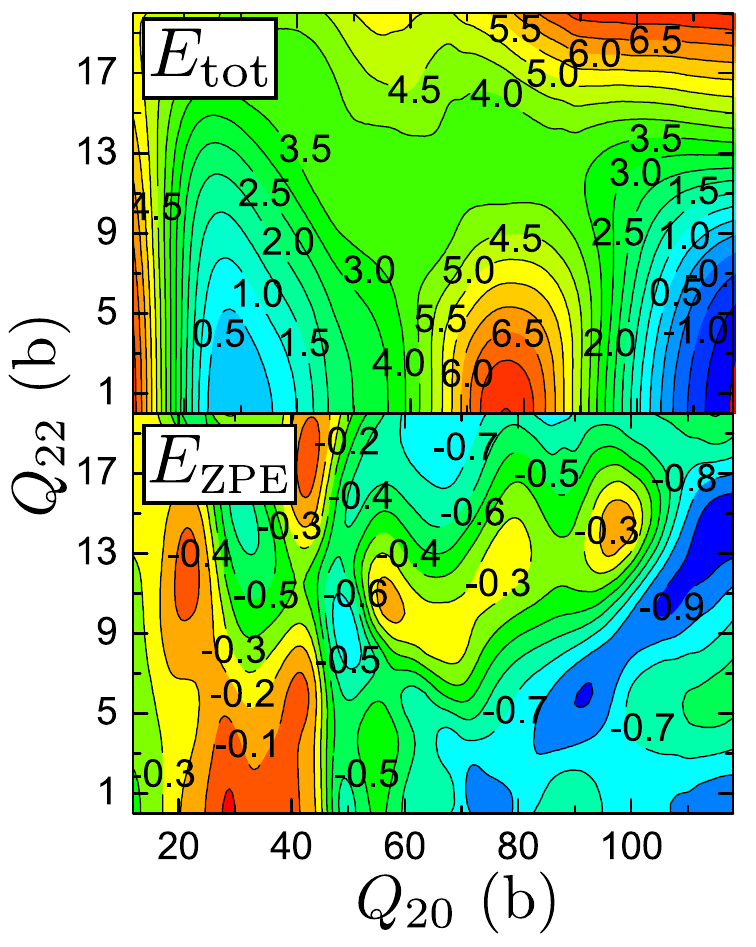}
\caption[C1]{\label{plot1}
(Color online) Contour plots of $E_{\text{tot}}$ (top) and $E_{\text{ZPE}}$  (bottom), in MeV, calculated
for $^{264}$Fm in the ($Q_{20}, Q_{22}$) collective space.
Both energies are plotted relatively to the
ground-state values.
}
\end{figure}

\textit{Results.}---The energies  $E_{\text{tot}}$ and $E_{\text{ZPE}}$
for $^{264}$Fm are shown in Fig.~\ref{plot1}. It is seen that in the range of deformations considered, the zero-point energy
varies between 0 and -0.9\,MeV; hence, it only slightly renormalizes the
topography of $E_{\text{tot}}$. The triaxial deformations are important around the fission barrier, and they reduce the fission
barrier height by over 2\,MeV.

\begin{figure}[htb]
\includegraphics[width=1.0\columnwidth]{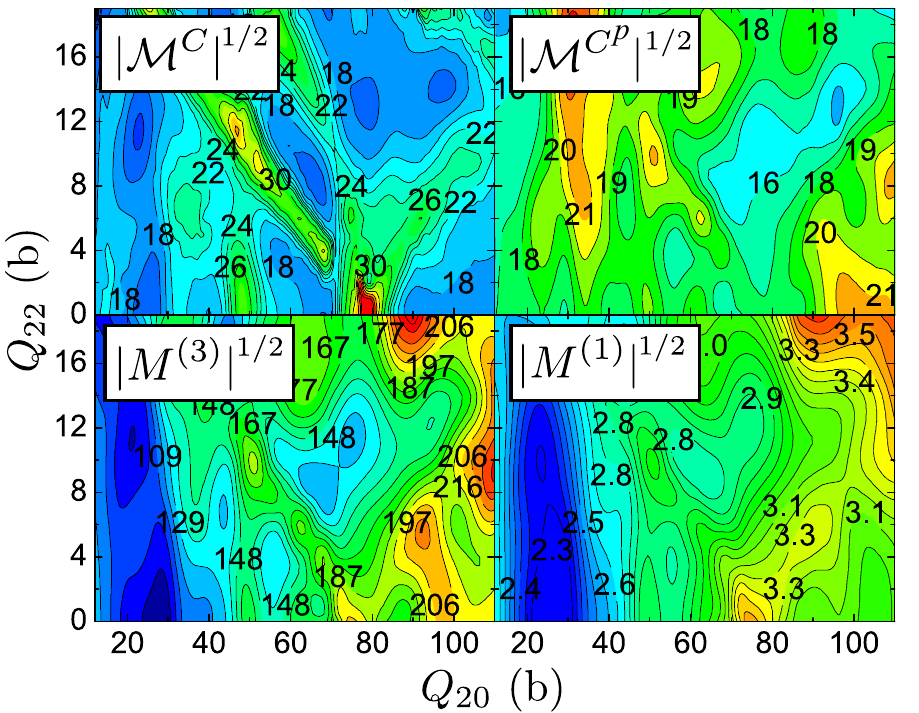}
\caption[C1]{\label{plot3}
(Color online)
Top: Square-root-determinants of inertia tensors $|\mathcal{M}^C|^{1/2}$ and
$|\mathcal{M}^{C^{\rm p}}|^{1/2}$ (in  $\hbar^2$\,MeV$^{-1}$\,b$^{-2}$/1000).
Bottom: energy-weighted moment tensors $|M^{(3)}|^{1/2}$ (in b$^2$\,MeV$^{-3}$/1000) and
$|M^{(1)}|^{1/2}$ (in b$^2$/MeV).
}
\end{figure}

Since the three individual components of the quadrupole
inertia tensor are difficult to interpret,
 in Fig.~\ref{plot3} we show
the square-root-determinants of inertia tensors (\ref{cranking-mass}) and (\ref{pc-mass}), defined as
$|\mathcal{M}|^{1/2}=(\mathcal{M}_{11}\mathcal{M}_{22}-\mathcal{M}_{12}^2)^{1/2}$.
These quantities are invariant with respect to two-dimensional rotations in the
space of collective
coordinates and well illustrate the overall magnitudes of collective
masses.
Figure~\ref{plot3} also shows the
square-root-determinants of energy-weighted moment tensors $M^{(1)}$
and $M^{(3)}$ (\ref{moment-tensors}) that define the
perturbative-cranking inertia tensor $\mathcal{M}^{C^{\rm p}}$.
For $|\mathcal{M}^C|^{1/2}$, we notice rapid variations as a function of
collective coordinates, and similar trends are
also evident for $|M^{(3)}|^{1/2}$ and $|M^{(1)}|^{1/2}$. However, in $\mathcal{M}^{C^{\rm
p}}$, which depends on the ratio (\ref{pc-mass}) of $M^{(3)}$ and $M^{(1)}$,
variations in  ($Q_{20}, Q_{22}$) are quenched.

\begin{figure}[htb]
\includegraphics[width=1.0\columnwidth]{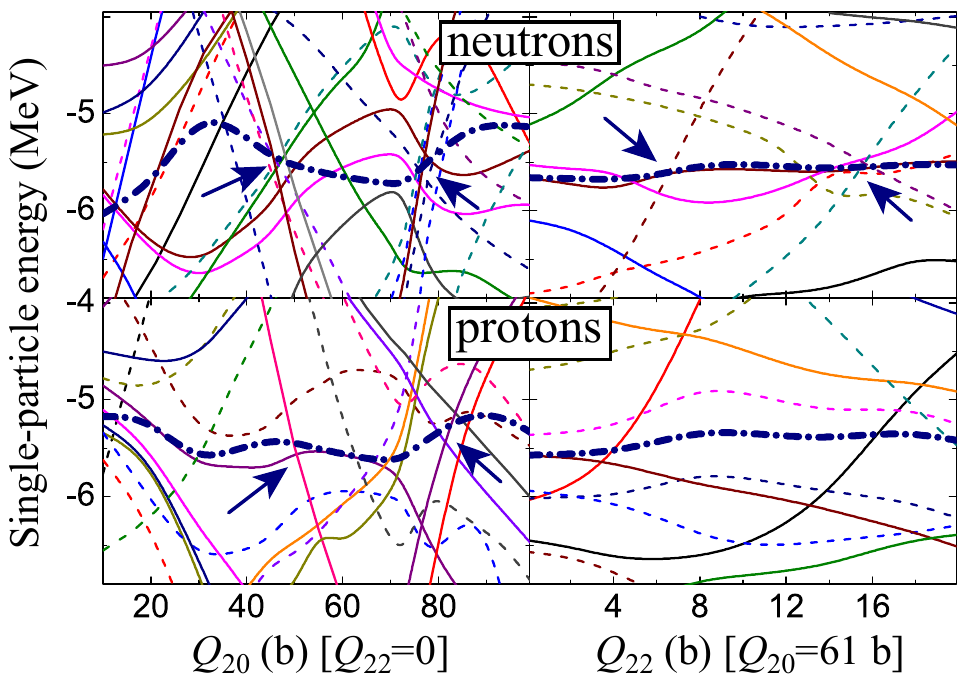}
\caption[C1]{\label{plot4}
(Color online) Single neutron (top) and proton (bottom)  energies for $^{264}$Fm as functions of
$Q_{20}$ (left, at $Q_{22}=0$) and $Q_{22}$ (right, at $Q_{20}=61$\,b). Thick dash-dotted lines showmark
Fermi energies. The arrows point to the level crossing regions discussed in the
text.}
\end{figure}
Large fluctuations of mass parameters are
manifestations of crossings of single-particle levels at the Fermi
level, see, e.g., Ref.~\cite{Nazarewicz1993489}. To illustrate this,
Fig.~\ref{plot4} displays  single-particle energies for $^{264}$Fm  along two straight lines in the collective
space, given by $Q_{22}=0$ and $Q_{20}=61$\,b. It is clearly visible,
that multiple level crossings appear very close to the Fermi energy,
at deformations where $\mathcal{M}^C$, $M^{(1)}$,  and $M^{(3)}$  exhibit rapid variations.

\begin{figure}[htb]
\includegraphics[width=1.0\columnwidth]{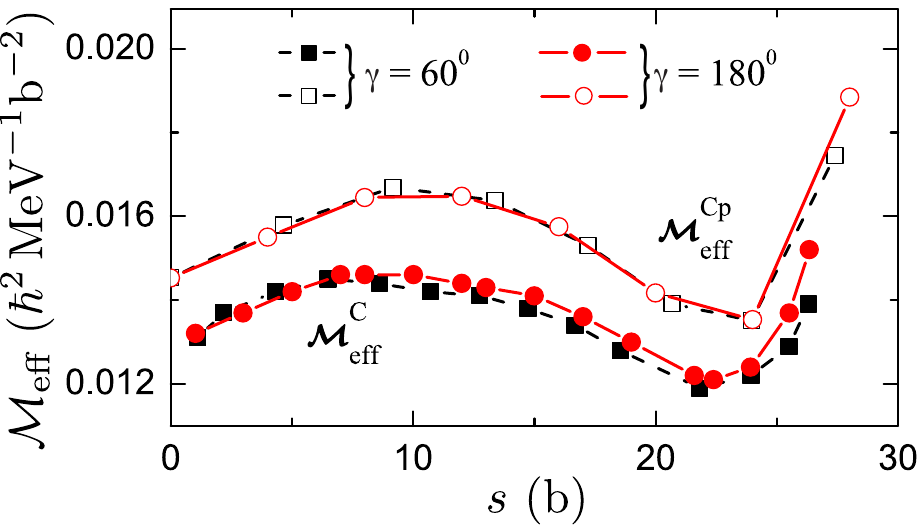}
\caption[C1]{\label{plot5}
(Color online) Effective quadrupole inertia $M_{\text{eff}}^{C}$
(full symbols) and $M_{\text{eff}}^{C^{\rm p}}$ (open symbols) as a
function of the path's length $s$, calculated for $\gamma=180^\circ$
(circles) and $60^\circ$ (squares).
}
\end{figure}
Before we proceed, we demonstrate the numerical
accuracy of the calculated inertia tensors. To this end,
we compute the effective quadrupole
inertia (\ref{eff-mass}) along the negative $Q_{20}$ axis, which corresponds to
$\gamma=180^\circ$. In this case, the $z$-axis is a symmetry axis and nuclear shape has an oblate deformation. Here, only $\mathcal{M}_{11}$ component
contributes to $\mathcal{M}_{\text{eff}}$ as $dq_2/ds=0$.
Next, nuclear densities were rotated in space,
so that the $y$-axis becomes the symmetry axis and $\gamma=60^\circ$. Along this path, all $\mathcal{M}_{ij}$ components are nonzero .
Nevertheless, the new path proceeds through exactly the same sequence
of shapes; hence, the effective quadrupole inertia must be
identical in both cases. This is demonstrated in Fig.~\ref{plot5}: the agreement between $\gamma=180^\circ$ and $\gamma=60^\circ$ results is indeed excellent.

We determined the minimum-action
paths by following two different numerical techniques: the
dynamic-programming method (DPM)~\cite{Baran198183}
and Ritz method (RM)~\cite{Baran19788}.
For DPM, we discretized the quadrupole surface into a two-dimensional mesh. Then, the fission path is calculated by connecting those mesh points that contribute minimum value for $S(L)$ out of all possible combinations. In case of comparatively large distances between two successive mesh points, we further divided the corresponding path length into small segments. This is essential in case of non-perturbative inertia as $\mathcal{M}^C$ varies quite nonlinearly in certain regions of the deformation plane.
In case of RM, trial
paths are expressed as Fourier series of collective coordinates and
the coefficients of different Fourier components are then extracted
by minimizing the action integral (\ref{action-integral}).
For both methods, different possible values of turning points  $s_{\rm in}$ and $s_{\rm out}$ have been considered to obtain the minimum action path.

\begin{figure}[htb]
\includegraphics[width=1.0\columnwidth]{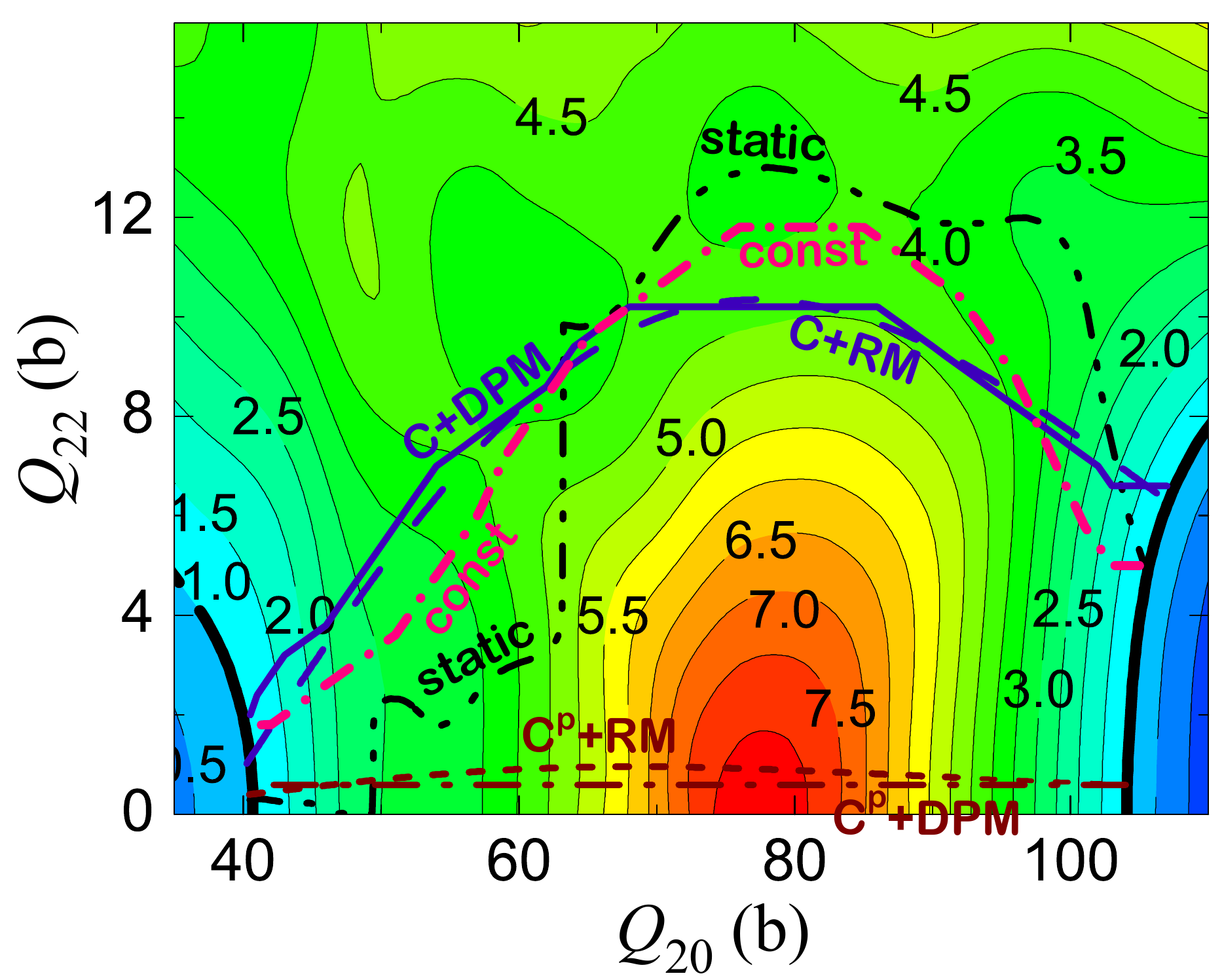}
\caption[C1]
{\label{plot6}
(Color online) Dynamic paths for spontaneous
fission of $^{264}$Fm, calculated for the non-perturbative $\mathcal{M}^C$
and perturbative $\mathcal{M}^{C^{\rm p}}$ cranking inertia using DMP and RM techniques to minimize the collective action integral (\ref{action-integral}).
The static pathway (``static") and that corresponding to a constant inertia (``const") are also shown.
The trajectories of turning points $s_{\rm in}$ and $s_{\rm out}$ are marked by thick solid lines.
}
\end{figure}
In Fig.~\ref{plot6} we show dynamical minimum-action paths determined
for $\mathcal{M}^C$ and $\mathcal{M}^{C^{\rm p}}$ and compare them
with the static path corresponding to the minimized collective
potential. To obtain the static path, we proceed from one point to the next point on the path
by searching the minimum potential on the circumference of a tiny circle centered around the previous point.
Evidently,
the static path traverses the longest distance through the
two-dimensional collective space.
For perturbative inertia,  there is a strong dynamical hindrance that prevents the paths
from departing towards large triaxial shapes: the collective mass
$\mathcal{M}^{C^{\rm p}}$ favors near-axial shapes.
This observation is consistent with findings of previous Refs.~\cite{Smo93,Smo95,Ghe99,Del06,War12}  that the effects of triaxial shapes on the fission
process are weakened by the inertia tensor. In the barrier region, all the components of $\mathcal{M}^{C^{\rm p}}$ increase smoothly with $Q_{22}$,  and this offsets the reduction of $V_{\text{eff}}$ in the action integral (\ref{action-integral}). Consequently, as discussed in Ref.~\cite{Ghe99}, the fission path remains fairly straight in order to achieve minimum action by minimizing the path's length.

This is not true for the non-perturbative inertia. Here,
localized large variations in $\mathcal{M}^C$   due to level crossings push the fission path substantially towards triaxial
deformations. Interestingly, the resulting non-perturbative trajectory appears fairly close to the static fission valley. Both of these trajectories indeed try to minimize the single particle level density along the fission path by avoiding the regions of level crossing shown in Fig.~\ref{plot3}.
For completeness,
the dynamical path corresponding to constant mass parameter (``const"), fixed
at the ground state value of $\mathcal{M}^C_{11}$, is shown in Fig.~\ref{plot6}. As expected, this  path also appears  close to the static fission valley.
Values of the action
integral and fission half-lives corresponding to different fission
paths are summarized in Table~\ref{table1}. They indicate strong structural dependence of the
spontaneous fission on collective dynamics. (It is worth emphasizing  that in the present theoretical work, the values of $T_{1/2}$ are calculated only for a comparative study and they are not intended to relate to experimental systematics.)
\begin{table}[htb]
\caption[C1]
{\label{table1}
Values of the action integral~(\protect\ref{action-integral}) and
half-lives for different spontaneous
fission paths shown in Fig.~\protect\ref{plot6}.
}
\begin{ruledtabular}
\begin{tabular}{llr}
path        &$S(L)$&$\log(T_{1/2}$/yr)\\
\hline
Static+$\mathcal{M}^C$&   23.4  &   -7.7  \\
Static+$\mathcal{M}^{C^{\rm p}}$& 20.8    &   -10.0\\
DPM+$\mathcal{M}^C$&  19.1    &  -11.4  \\
RM+$\mathcal{M}^C$&  18.9   &   -11.6  \\
DPM+$\mathcal{M}^{C^{\rm p}}$&  16.8  &  -13.4  \\
RM+$\mathcal{M}^{C^{\rm p}}$&  16.8   &   -13.4  \\
\end{tabular}
\end{ruledtabular}
\end{table}

\textit{Conclusions.}---In conclusion, SF lifetimes have been studied within
a dynamic approach based on the minimization of the fission action
in a two-dimensional quadrupole collective space of elongation and triaxiality.
Strong dynamical effects have been predicted.
In the perturbative picture, the collective inertia $\mathcal{M}^{C^{\rm p}}$ drives the system towards near-axial shapes, consistent with
Refs.~\cite{Smo93,Smo95,Ghe99,Del06,War12}. We believe that this effect is an artifact of the perturbative approximation employed that underestimates the role of level crossings.
The most important conclusion of this study is that, strong triaxial effects are
indeed predicted with the more appropriate non-perturbative
cranking inertia, in agreement with static calculations. This indicates that the
localized structural properties of collective mass parameter, present
in $\mathcal{M}^{C}$, play a crucial role in determining the SF dynamics. Presently, the inertia tensors are calculated within the cranking variant of ATDHFB  \cite{Bar11}. The calculation of the full ATDHFB inertia is in progress, also developments are initiated in the context of
dynamical effects due to the competition between triaxial and reflection asymmetric degrees
of freedom.

\begin{acknowledgments}
Useful discussions with L.\ Pr\'ochniak and A. Staszczak are
gratefully acknowledged. The paper is finalized at the INT Program INT-13-3 in Seattle, USA. This work was supported by the U.S.\
Department of Energy under Contracts No.\ DE-FG02-96ER40963
(University of Tennessee), No. DE-FG52-09NA29461 (the Stewardship
Science Academic Alliances program), and No.\ DE-SC0008499    (NUCLEI
SciDAC Collaboration), by the Academy of Finland and University
of Jyv\"askyl\"a within the FIDIPRO programme, and by the
Polish National Science Center under Contract No.~2012/07/B/ST2/03907 and No.~2011/01/B/ST2/03667.
An award of computer
time was provided by the National Institute for Computational
Sciences (NICS) and the Innovative and Novel Computational Impact on
Theory and Experiment (INCITE) program using resources of the OLCF facility.
\end{acknowledgments}

\bibliography{ref}

\begin{thebibliography}{36}%
\makeatletter
\providecommand \@ifxundefined [1]{%
 \@ifx{#1\undefined}
}%
\providecommand \@ifnum [1]{%
 \ifnum #1\expandafter \@firstoftwo
 \else \expandafter \@secondoftwo
 \fi
}%
\providecommand \@ifx [1]{%
 \ifx #1\expandafter \@firstoftwo
 \else \expandafter \@secondoftwo
 \fi
}%
\providecommand \natexlab [1]{#1}%
\providecommand \enquote  [1]{``#1''}%
\providecommand \bibnamefont  [1]{#1}%
\providecommand \bibfnamefont [1]{#1}%
\providecommand \citenamefont [1]{#1}%
\providecommand \href@noop [0]{\@secondoftwo}%
\providecommand \href [0]{\begingroup \@sanitize@url \@href}%
\providecommand \@href[1]{\@@startlink{#1}\@@href}%
\providecommand \@@href[1]{\endgroup#1\@@endlink}%
\providecommand \@sanitize@url [0]{\catcode `\\12\catcode `\$12\catcode
  `\&12\catcode `\#12\catcode `\^12\catcode `\_12\catcode `\%12\relax}%
\providecommand \@@startlink[1]{}%
\providecommand \@@endlink[0]{}%
\providecommand \url  [0]{\begingroup\@sanitize@url \@url }%
\providecommand \@url [1]{\endgroup\@href {#1}{\urlprefix }}%
\providecommand \urlprefix  [0]{URL }%
\providecommand \Eprint [0]{\href }%
\providecommand \doibase [0]{http://dx.doi.org/}%
\providecommand \selectlanguage [0]{\@gobble}%
\providecommand \bibinfo  [0]{\@secondoftwo}%
\providecommand \bibfield  [0]{\@secondoftwo}%
\providecommand \translation [1]{[#1]}%
\providecommand \BibitemOpen [0]{}%
\providecommand \bibitemStop [0]{}%
\providecommand \bibitemNoStop [0]{.\EOS\space}%
\providecommand \EOS [0]{\spacefactor3000\relax}%
\providecommand \BibitemShut  [1]{\csname bibitem#1\endcsname}%
\let\auto@bib@innerbib\@empty
\bibitem [{\citenamefont {Wagemans}(1991)}]{Wagemans}%
  \BibitemOpen
  \bibfield  {author} {\bibinfo {author} {\bibfnamefont {C.}~\bibnamefont
  {Wagemans}},\ }\href@noop {} {\emph {\bibinfo {title} {The Nuclear Fission
  Process}}}\ (\bibinfo  {publisher} {CRC Press, Boca Raton},\ \bibinfo {year}
  {1991})\BibitemShut {NoStop}%
\bibitem [{\citenamefont {Krappe}\ and\ \citenamefont
  {Pomorski}(2012)}]{KrappePom}%
  \BibitemOpen
  \bibfield  {author} {\bibinfo {author} {\bibfnamefont {H.~J.}\ \bibnamefont
  {Krappe}}\ and\ \bibinfo {author} {\bibfnamefont {K.}~\bibnamefont
  {Pomorski}},\ }\href@noop {} {\emph {\bibinfo {title} {Theory of Nuclear
  Fission: A Textbook}}}\ (\bibinfo  {publisher} {Springer},\ \bibinfo
  {address} {New York},\ \bibinfo {year} {2012})\BibitemShut {NoStop}%
\bibitem [{\citenamefont {Shultis}\ and\ \citenamefont
  {Faw}(2007)}]{ShultisFaw}%
  \BibitemOpen
  \bibfield  {author} {\bibinfo {author} {\bibfnamefont {J.~K.}\ \bibnamefont
  {Shultis}}\ and\ \bibinfo {author} {\bibfnamefont {R.~E.}\ \bibnamefont
  {Faw}},\ }\href@noop {} {\emph {\bibinfo {title} {Fundamentals of Nuclear
  Science and Engineering Second Edition}}}\ (\bibinfo  {publisher} {CRC Press,
  Boca Raton},\ \bibinfo {year} {2007})\BibitemShut {NoStop}%
\bibitem [{\citenamefont {Oganessian}(2007)}]{Oganessian07}%
  \BibitemOpen
  \bibfield  {author} {\bibinfo {author} {\bibfnamefont {Y.~T.}\ \bibnamefont
  {Oganessian}},\ }\href@noop {} {\bibfield  {journal} {\bibinfo  {journal} {J.
  Phys. G}\ }\textbf {\bibinfo {volume} {34}},\ \bibinfo {pages} {R165}
  (\bibinfo {year} {2007})}\BibitemShut {NoStop}%
\bibitem [{\citenamefont {Staszczak}\ \emph {et~al.}(2013)\citenamefont
  {Staszczak}, \citenamefont {Baran},\ and\ \citenamefont
  {Nazarewicz}}]{Sta13}%
  \BibitemOpen
  \bibfield  {author} {\bibinfo {author} {\bibfnamefont {A.}~\bibnamefont
  {Staszczak}}, \bibinfo {author} {\bibfnamefont {A.}~\bibnamefont {Baran}}, \
  and\ \bibinfo {author} {\bibfnamefont {W.}~\bibnamefont {Nazarewicz}},\
  }\href@noop {} {\bibfield  {journal} {\bibinfo  {journal} {Phys. Rev. C}\
  }\textbf {\bibinfo {volume} {87}},\ \bibinfo {pages} {024320} (\bibinfo
  {year} {2013})}\BibitemShut {NoStop}%
\bibitem [{\citenamefont {Arnould}\ \emph {et~al.}(2007)\citenamefont
  {Arnould}, \citenamefont {Goriely},\ and\ \citenamefont {Takahashi}}]{Arn07}%
  \BibitemOpen
  \bibfield  {author} {\bibinfo {author} {\bibfnamefont {M.}~\bibnamefont
  {Arnould}}, \bibinfo {author} {\bibfnamefont {S.}~\bibnamefont {Goriely}}, \
  and\ \bibinfo {author} {\bibfnamefont {K.}~\bibnamefont {Takahashi}},\
  }\href@noop {} {\bibfield  {journal} {\bibinfo  {journal} {Phys. Rep.}\
  }\textbf {\bibinfo {volume} {450}},\ \bibinfo {pages} {97} (\bibinfo {year}
  {2007})}\BibitemShut {NoStop}%
\bibitem [{\citenamefont {Panov}\ \emph {et~al.}(2010)\citenamefont {Panov},
  \citenamefont {Korneev}, \citenamefont {Rauscher}, \citenamefont
  {Mart\'{\i}nez-Pinedo}, \citenamefont {Kelic-Heil}, \citenamefont {Zinner},\
  and\ \citenamefont {Thielemann}}]{Pan10}%
  \BibitemOpen
  \bibfield  {author} {\bibinfo {author} {\bibfnamefont {I.~V.}\ \bibnamefont
  {Panov}}, \bibinfo {author} {\bibfnamefont {I.~Y.}\ \bibnamefont {Korneev}},
  \bibinfo {author} {\bibfnamefont {T.}~\bibnamefont {Rauscher}}, \bibinfo
  {author} {\bibfnamefont {G.}~\bibnamefont {Mart\'{\i}nez-Pinedo}}, \bibinfo
  {author} {\bibfnamefont {A.}~\bibnamefont {Kelic-Heil}}, \bibinfo {author}
  {\bibfnamefont {N.~T.}\ \bibnamefont {Zinner}}, \ and\ \bibinfo {author}
  {\bibfnamefont {F.-K.}\ \bibnamefont {Thielemann}},\ }\href@noop {}
  {\bibfield  {journal} {\bibinfo  {journal} {A\&A}\ }\textbf {\bibinfo
  {volume} {513}},\ \bibinfo {pages} {A61} (\bibinfo {year}
  {2010})}\BibitemShut {NoStop}%
\bibitem [{\citenamefont {Erler}\ \emph {et~al.}(2012)\citenamefont {Erler},
  \citenamefont {Langanke}, \citenamefont {Loens}, \citenamefont
  {Martinez-Pinedo},\ and\ \citenamefont {Reinhard}}]{ErlLan12}%
  \BibitemOpen
  \bibfield  {author} {\bibinfo {author} {\bibfnamefont {J.}~\bibnamefont
  {Erler}}, \bibinfo {author} {\bibfnamefont {K.}~\bibnamefont {Langanke}},
  \bibinfo {author} {\bibfnamefont {H.~P.}\ \bibnamefont {Loens}}, \bibinfo
  {author} {\bibfnamefont {G.}~\bibnamefont {Martinez-Pinedo}}, \ and\ \bibinfo
  {author} {\bibfnamefont {P.-G.}\ \bibnamefont {Reinhard}},\ }\href@noop {}
  {\bibfield  {journal} {\bibinfo  {journal} {Phys. Rev. C}\ }\textbf {\bibinfo
  {volume} {85}},\ \bibinfo {pages} {025802} (\bibinfo {year}
  {2012})}\BibitemShut {NoStop}%
\bibitem [{\citenamefont {Bender}\ \emph {et~al.}(2003)\citenamefont {Bender},
  \citenamefont {Heenen},\ and\ \citenamefont {Reinhard}}]{Ben03}%
  \BibitemOpen
  \bibfield  {author} {\bibinfo {author} {\bibfnamefont {M.}~\bibnamefont
  {Bender}}, \bibinfo {author} {\bibfnamefont {P.-H.}\ \bibnamefont {Heenen}},
  \ and\ \bibinfo {author} {\bibfnamefont {P.-G.}\ \bibnamefont {Reinhard}},\
  }\href@noop {} {\bibfield  {journal} {\bibinfo  {journal} {Rev. Mod. Phys.}\
  }\textbf {\bibinfo {volume} {75}},\ \bibinfo {pages} {121} (\bibinfo {year}
  {2003})}\BibitemShut {NoStop}%
\bibitem [{\citenamefont {Baranger}\ and\ \citenamefont
  {V{\'e}n{\'e}roni}(1978)}]{Baranger1978123}%
  \BibitemOpen
  \bibfield  {author} {\bibinfo {author} {\bibfnamefont {M.}~\bibnamefont
  {Baranger}}\ and\ \bibinfo {author} {\bibfnamefont {M.}~\bibnamefont
  {V{\'e}n{\'e}roni}},\ }\href@noop {} {\bibfield  {journal} {\bibinfo
  {journal} {Ann. Phys.}\ }\textbf {\bibinfo {volume} {114}},\ \bibinfo {pages}
  {123 } (\bibinfo {year} {1978})}\BibitemShut {NoStop}%
\bibitem [{\citenamefont {Dobaczewski}\ and\ \citenamefont
  {Skalski}(1981)}]{Doba81}%
  \BibitemOpen
  \bibfield  {author} {\bibinfo {author} {\bibfnamefont {J.}~\bibnamefont
  {Dobaczewski}}\ and\ \bibinfo {author} {\bibfnamefont {J.}~\bibnamefont
  {Skalski}},\ }\href@noop {} {\bibfield  {journal} {\bibinfo  {journal} {Nucl.
  Phys. A}\ }\textbf {\bibinfo {volume} {369}},\ \bibinfo {pages} {123 }
  (\bibinfo {year} {1981})}\BibitemShut {NoStop}%
\bibitem [{\citenamefont {Baran}\ \emph {et~al.}(2011)\citenamefont {Baran},
  \citenamefont {Sheikh}, \citenamefont {Dobaczewski}, \citenamefont
  {Nazarewicz},\ and\ \citenamefont {Staszczak}}]{Bar11}%
  \BibitemOpen
  \bibfield  {author} {\bibinfo {author} {\bibfnamefont {A.}~\bibnamefont
  {Baran}}, \bibinfo {author} {\bibfnamefont {J.~A.}\ \bibnamefont {Sheikh}},
  \bibinfo {author} {\bibfnamefont {J.}~\bibnamefont {Dobaczewski}}, \bibinfo
  {author} {\bibfnamefont {W.}~\bibnamefont {Nazarewicz}}, \ and\ \bibinfo
  {author} {\bibfnamefont {A.}~\bibnamefont {Staszczak}},\ }\href@noop {}
  {\bibfield  {journal} {\bibinfo  {journal} {Phys. Rev. C}\ }\textbf {\bibinfo
  {volume} {84}},\ \bibinfo {pages} {054321} (\bibinfo {year}
  {2011})}\BibitemShut {NoStop}%
\bibitem [{\citenamefont {Baran}\ \emph {et~al.}(1981)\citenamefont {Baran},
  \citenamefont {Pomorski}, \citenamefont {Lukasiak},\ and\ \citenamefont
  {Sobiczewski}}]{Baran198183}%
  \BibitemOpen
  \bibfield  {author} {\bibinfo {author} {\bibfnamefont {A.}~\bibnamefont
  {Baran}}, \bibinfo {author} {\bibfnamefont {K.}~\bibnamefont {Pomorski}},
  \bibinfo {author} {\bibfnamefont {A.}~\bibnamefont {Lukasiak}}, \ and\
  \bibinfo {author} {\bibfnamefont {A.}~\bibnamefont {Sobiczewski}},\ }\href
  {\doibase http://dx.doi.org/10.1016/0375-9474(81)90471-1} {\bibfield
  {journal} {\bibinfo  {journal} {Nucl. Phys. A}\ }\textbf {\bibinfo {volume}
  {361}},\ \bibinfo {pages} {83 } (\bibinfo {year} {1981})}\BibitemShut
  {NoStop}%
\bibitem [{\citenamefont {Baran}(1978)}]{Baran19788}%
  \BibitemOpen
  \bibfield  {author} {\bibinfo {author} {\bibfnamefont {A.}~\bibnamefont
  {Baran}},\ }\href {\doibase http://dx.doi.org/10.1016/0370-2693(78)90085-0}
  {\bibfield  {journal} {\bibinfo  {journal} {Phys. Lett. B}\ }\textbf
  {\bibinfo {volume} {76}},\ \bibinfo {pages} {8 } (\bibinfo {year}
  {1978})}\BibitemShut {NoStop}%
\bibitem [{\citenamefont {Brack}\ \emph {et~al.}(1972)\citenamefont {Brack},
  \citenamefont {Damgaard}, \citenamefont {Jensen}, \citenamefont {Pauli},
  \citenamefont {Strutinsky},\ and\ \citenamefont {Wong}}]{Bra72}%
  \BibitemOpen
  \bibfield  {author} {\bibinfo {author} {\bibfnamefont {M.}~\bibnamefont
  {Brack}}, \bibinfo {author} {\bibfnamefont {J.}~\bibnamefont {Damgaard}},
  \bibinfo {author} {\bibfnamefont {A.~S.}\ \bibnamefont {Jensen}}, \bibinfo
  {author} {\bibfnamefont {H.~C.}\ \bibnamefont {Pauli}}, \bibinfo {author}
  {\bibfnamefont {V.~M.}\ \bibnamefont {Strutinsky}}, \ and\ \bibinfo {author}
  {\bibfnamefont {C.~Y.}\ \bibnamefont {Wong}},\ }\href@noop {} {\bibfield
  {journal} {\bibinfo  {journal} {Rev. Mod. Phys.}\ }\textbf {\bibinfo {volume}
  {44}},\ \bibinfo {pages} {320} (\bibinfo {year} {1972})}\BibitemShut
  {NoStop}%
\bibitem [{\citenamefont {Skalski}(2008)}]{Ska08}%
  \BibitemOpen
  \bibfield  {author} {\bibinfo {author} {\bibfnamefont {J.}~\bibnamefont
  {Skalski}},\ }\href@noop {} {\bibfield  {journal} {\bibinfo  {journal} {Phys.
  Rev. C}\ }\textbf {\bibinfo {volume} {77}},\ \bibinfo {pages} {064610}
  (\bibinfo {year} {2008})}\BibitemShut {NoStop}%
\bibitem [{\citenamefont {Baran}\ \emph {et~al.}(2005)\citenamefont {Baran},
  \citenamefont {{\L}ojewski}, \citenamefont {Sieja},\ and\ \citenamefont
  {Kowal}}]{Baran05}%
  \BibitemOpen
  \bibfield  {author} {\bibinfo {author} {\bibfnamefont {A.}~\bibnamefont
  {Baran}}, \bibinfo {author} {\bibfnamefont {Z.}~\bibnamefont {{\L}ojewski}},
  \bibinfo {author} {\bibfnamefont {K.}~\bibnamefont {Sieja}}, \ and\ \bibinfo
  {author} {\bibfnamefont {M.}~\bibnamefont {Kowal}},\ }\href@noop {}
  {\bibfield  {journal} {\bibinfo  {journal} {Phys. Rev. C}\ }\textbf {\bibinfo
  {volume} {72}},\ \bibinfo {pages} {044310} (\bibinfo {year}
  {2005})}\BibitemShut {NoStop}%
\bibitem [{\citenamefont {Staszczak}\ \emph {et~al.}(2009)\citenamefont
  {Staszczak}, \citenamefont {Baran}, \citenamefont {Dobaczewski},\ and\
  \citenamefont {Nazarewicz}}]{Sta09}%
  \BibitemOpen
  \bibfield  {author} {\bibinfo {author} {\bibfnamefont {A.}~\bibnamefont
  {Staszczak}}, \bibinfo {author} {\bibfnamefont {A.}~\bibnamefont {Baran}},
  \bibinfo {author} {\bibfnamefont {J.}~\bibnamefont {Dobaczewski}}, \ and\
  \bibinfo {author} {\bibfnamefont {W.}~\bibnamefont {Nazarewicz}},\
  }\href@noop {} {\bibfield  {journal} {\bibinfo  {journal} {Phys. Rev. C}\
  }\textbf {\bibinfo {volume} {80}},\ \bibinfo {pages} {014309} (\bibinfo
  {year} {2009})}\BibitemShut {NoStop}%
\bibitem [{\citenamefont {Dobaczewski}\ and\ \citenamefont
  {Olbratowski}(2004)}]{doba04}%
  \BibitemOpen
  \bibfield  {author} {\bibinfo {author} {\bibfnamefont {J.}~\bibnamefont
  {Dobaczewski}}\ and\ \bibinfo {author} {\bibfnamefont {P.}~\bibnamefont
  {Olbratowski}},\ }\href@noop {} {\bibfield  {journal} {\bibinfo  {journal}
  {Comput.~Phys.~Commun.}\ }\textbf {\bibinfo {volume} {158}},\ \bibinfo
  {pages} {158} (\bibinfo {year} {2004})}\BibitemShut {NoStop}%
\bibitem [{\citenamefont {Schunck}\ \emph {et~al.}(2012)\citenamefont
  {Schunck}, \citenamefont {Dobaczewski}, \citenamefont {McDonnell},
  \citenamefont {Satu{\l}a}, \citenamefont {Sheikh}, \citenamefont {Staszczak},
  \citenamefont {Stoitsov},\ and\ \citenamefont {Toivanen}}]{Schunck2012166}%
  \BibitemOpen
  \bibfield  {author} {\bibinfo {author} {\bibfnamefont {N.}~\bibnamefont
  {Schunck}}, \bibinfo {author} {\bibfnamefont {J.}~\bibnamefont
  {Dobaczewski}}, \bibinfo {author} {\bibfnamefont {J.}~\bibnamefont
  {McDonnell}}, \bibinfo {author} {\bibfnamefont {W.}~\bibnamefont
  {Satu{\l}a}}, \bibinfo {author} {\bibfnamefont {J.}~\bibnamefont {Sheikh}},
  \bibinfo {author} {\bibfnamefont {A.}~\bibnamefont {Staszczak}}, \bibinfo
  {author} {\bibfnamefont {M.}~\bibnamefont {Stoitsov}}, \ and\ \bibinfo
  {author} {\bibfnamefont {P.}~\bibnamefont {Toivanen}},\ }\href@noop {}
  {\bibfield  {journal} {\bibinfo  {journal} {Comput. Phys. Commun.}\ }\textbf
  {\bibinfo {volume} {183}},\ \bibinfo {pages} {166} (\bibinfo {year}
  {2012})}\BibitemShut {NoStop}%
\bibitem [{\citenamefont {Bartel}\ \emph {et~al.}(1982)\citenamefont {Bartel},
  \citenamefont {Quentin}, \citenamefont {Brack}, \citenamefont {Guet},\ and\
  \citenamefont {Håkansson}}]{Bartel198279}%
  \BibitemOpen
  \bibfield  {author} {\bibinfo {author} {\bibfnamefont {J.}~\bibnamefont
  {Bartel}}, \bibinfo {author} {\bibfnamefont {P.}~\bibnamefont {Quentin}},
  \bibinfo {author} {\bibfnamefont {M.}~\bibnamefont {Brack}}, \bibinfo
  {author} {\bibfnamefont {C.}~\bibnamefont {Guet}}, \ and\ \bibinfo {author}
  {\bibfnamefont {H.-B.}\ \bibnamefont {Håkansson}},\ }\href {\doibase
  http://dx.doi.org/10.1016/0375-9474(82)90403-1} {\bibfield  {journal}
  {\bibinfo  {journal} {Nucl. Phys. A}\ }\textbf {\bibinfo {volume} {386}},\
  \bibinfo {pages} {79 } (\bibinfo {year} {1982})}\BibitemShut {NoStop}%
\bibitem [{\citenamefont {Dobaczewski}\ \emph {et~al.}(2002)\citenamefont
  {Dobaczewski}, \citenamefont {Nazarewicz},\ and\ \citenamefont
  {Stoitsov}}]{doba02}%
  \BibitemOpen
  \bibfield  {author} {\bibinfo {author} {\bibfnamefont {J.}~\bibnamefont
  {Dobaczewski}}, \bibinfo {author} {\bibfnamefont {W.}~\bibnamefont
  {Nazarewicz}}, \ and\ \bibinfo {author} {\bibfnamefont {M.}~\bibnamefont
  {Stoitsov}},\ }\href@noop {} {\bibfield  {journal} {\bibinfo  {journal} {Eur.
  Phys. J. A}\ }\textbf {\bibinfo {volume} {15}},\ \bibinfo {pages} {21}
  (\bibinfo {year} {2002})}\BibitemShut {NoStop}%
\bibitem [{\citenamefont {Staszczak}\ \emph {et~al.}(1989)\citenamefont
  {Staszczak}, \citenamefont {Pi{\l}at},\ and\ \citenamefont
  {Pomorski}}]{Staszczak1989589}%
  \BibitemOpen
  \bibfield  {author} {\bibinfo {author} {\bibfnamefont {A.}~\bibnamefont
  {Staszczak}}, \bibinfo {author} {\bibfnamefont {S.}~\bibnamefont {Pi{\l}at}},
  \ and\ \bibinfo {author} {\bibfnamefont {K.}~\bibnamefont {Pomorski}},\
  }\href@noop {} {\bibfield  {journal} {\bibinfo  {journal} {Nucl. Phys. A}\
  }\textbf {\bibinfo {volume} {504}},\ \bibinfo {pages} {589} (\bibinfo {year}
  {1989})}\BibitemShut {NoStop}%
\bibitem [{\citenamefont {Baran}\ \emph {et~al.}(2007)\citenamefont {Baran},
  \citenamefont {Staszczak}, \citenamefont {Dobaczewski},\ and\ \citenamefont
  {Nazarewicz}}]{Bar07}%
  \BibitemOpen
  \bibfield  {author} {\bibinfo {author} {\bibfnamefont {A.}~\bibnamefont
  {Baran}}, \bibinfo {author} {\bibfnamefont {A.}~\bibnamefont {Staszczak}},
  \bibinfo {author} {\bibfnamefont {J.}~\bibnamefont {Dobaczewski}}, \ and\
  \bibinfo {author} {\bibfnamefont {W.}~\bibnamefont {Nazarewicz}},\
  }\href@noop {} {\bibfield  {journal} {\bibinfo  {journal} {Int. J. Mod. Phys.
  E}\ }\textbf {\bibinfo {volume} {16}},\ \bibinfo {pages} {443} (\bibinfo
  {year} {2007})}\BibitemShut {NoStop}%
\bibitem [{\citenamefont {Giannoni}\ and\ \citenamefont
  {Quentin}(1980)}]{Gia80}%
  \BibitemOpen
  \bibfield  {author} {\bibinfo {author} {\bibfnamefont {M.~J.}\ \bibnamefont
  {Giannoni}}\ and\ \bibinfo {author} {\bibfnamefont {P.}~\bibnamefont
  {Quentin}},\ }\href@noop {} {\bibfield  {journal} {\bibinfo  {journal} {Phys.
  Rev. C}\ }\textbf {\bibinfo {volume} {21}},\ \bibinfo {pages} {2076}
  (\bibinfo {year} {1980})}\BibitemShut {NoStop}%
\bibitem [{\citenamefont {Yuldashbaeva}\ \emph {et~al.}(1999)\citenamefont
  {Yuldashbaeva}, \citenamefont {Libert}, \citenamefont {Quentin},\ and\
  \citenamefont {Girod}}]{Yuldashbaeva19991}%
  \BibitemOpen
  \bibfield  {author} {\bibinfo {author} {\bibfnamefont {E.}~\bibnamefont
  {Yuldashbaeva}}, \bibinfo {author} {\bibfnamefont {J.}~\bibnamefont
  {Libert}}, \bibinfo {author} {\bibfnamefont {P.}~\bibnamefont {Quentin}}, \
  and\ \bibinfo {author} {\bibfnamefont {M.}~\bibnamefont {Girod}},\
  }\href@noop {} {\bibfield  {journal} {\bibinfo  {journal} {Phys. Lett. B}\
  }\textbf {\bibinfo {volume} {461}},\ \bibinfo {pages} {1} (\bibinfo {year}
  {1999})}\BibitemShut {NoStop}%
\bibitem [{Note1()}]{Note1}%
  \BibitemOpen
  \bibinfo {note} {Here we corrected misprinted equations (56) and (60) of
  Ref.~\cite {Bar11}.}\BibitemShut {Stop}%
\bibitem [{\citenamefont {Baldo}\ \emph {et~al.}(2013)\citenamefont {Baldo},
  \citenamefont {Robledo}, \citenamefont {Schuck},\ and\ \citenamefont
  {Vi\~nas}}]{Baldo13}%
  \BibitemOpen
  \bibfield  {author} {\bibinfo {author} {\bibfnamefont {M.}~\bibnamefont
  {Baldo}}, \bibinfo {author} {\bibfnamefont {L.~M.}\ \bibnamefont {Robledo}},
  \bibinfo {author} {\bibfnamefont {P.}~\bibnamefont {Schuck}}, \ and\ \bibinfo
  {author} {\bibfnamefont {X.}~\bibnamefont {Vi\~nas}},\ }\href@noop {}
  {\bibfield  {journal} {\bibinfo  {journal} {Phys. Rev. C}\ }\textbf {\bibinfo
  {volume} {87}},\ \bibinfo {pages} {064305} (\bibinfo {year}
  {2013})}\BibitemShut {NoStop}%
\bibitem [{\citenamefont {Giuliani}\ and\ \citenamefont
  {Robledo}(2013)}]{Giuliani13}%
  \BibitemOpen
  \bibfield  {author} {\bibinfo {author} {\bibfnamefont {S.~A.}\ \bibnamefont
  {Giuliani}}\ and\ \bibinfo {author} {\bibfnamefont {L.~M.}\ \bibnamefont
  {Robledo}},\ }\href@noop {} {\  (\bibinfo {year} {2013})},\ \Eprint
  {http://arxiv.org/abs/1305.0293} {arXiv:1305.0293 [nucl-th]} \BibitemShut
  {NoStop}%
\bibitem [{\citenamefont {Warda}\ \emph {et~al.}(2002)\citenamefont {Warda},
  \citenamefont {Egido}, \citenamefont {Robledo},\ and\ \citenamefont
  {Pomorski}}]{Warda02}%
  \BibitemOpen
  \bibfield  {author} {\bibinfo {author} {\bibfnamefont {M.}~\bibnamefont
  {Warda}}, \bibinfo {author} {\bibfnamefont {J.~L.}\ \bibnamefont {Egido}},
  \bibinfo {author} {\bibfnamefont {L.~M.}\ \bibnamefont {Robledo}}, \ and\
  \bibinfo {author} {\bibfnamefont {K.}~\bibnamefont {Pomorski}},\ }\href@noop
  {} {\bibfield  {journal} {\bibinfo  {journal} {Phys. Rev. C}\ }\textbf
  {\bibinfo {volume} {66}},\ \bibinfo {pages} {014310} (\bibinfo {year}
  {2002})}\BibitemShut {NoStop}%
\bibitem [{\citenamefont {Smola{\'n}czuk}\ \emph {et~al.}(1993)\citenamefont
  {Smola{\'n}czuk}, \citenamefont {Klapdor-Kleingrothaus},\ and\ \citenamefont
  {Sobiczewski}}]{Smo93}%
  \BibitemOpen
  \bibfield  {author} {\bibinfo {author} {\bibfnamefont {R.}~\bibnamefont
  {Smola{\'n}czuk}}, \bibinfo {author} {\bibfnamefont {H.}~\bibnamefont
  {Klapdor-Kleingrothaus}}, \ and\ \bibinfo {author} {\bibfnamefont
  {A.}~\bibnamefont {Sobiczewski}},\ }\href@noop {} {\bibfield  {journal}
  {\bibinfo  {journal} {Acta Phys. Pol. B}\ }\textbf {\bibinfo {volume} {24}},\
  \bibinfo {pages} {685} (\bibinfo {year} {1993})}\BibitemShut {NoStop}%
\bibitem [{\citenamefont {Smola{\'n}czuk}\ \emph {et~al.}(1995)\citenamefont
  {Smola{\'n}czuk}, \citenamefont {Skalski},\ and\ \citenamefont
  {Sobiczewski}}]{Smo95}%
  \BibitemOpen
  \bibfield  {author} {\bibinfo {author} {\bibfnamefont {R.}~\bibnamefont
  {Smola{\'n}czuk}}, \bibinfo {author} {\bibfnamefont {J.}~\bibnamefont
  {Skalski}}, \ and\ \bibinfo {author} {\bibfnamefont {A.}~\bibnamefont
  {Sobiczewski}},\ }\href@noop {} {\bibfield  {journal} {\bibinfo  {journal}
  {Phys. Rev. C}\ }\textbf {\bibinfo {volume} {52}},\ \bibinfo {pages} {1871}
  (\bibinfo {year} {1995})}\BibitemShut {NoStop}%
\bibitem [{\citenamefont {Gherghescu}\ \emph {et~al.}(1999)\citenamefont
  {Gherghescu}, \citenamefont {Skalski}, \citenamefont {Patyk},\ and\
  \citenamefont {Sobiczewski}}]{Ghe99}%
  \BibitemOpen
  \bibfield  {author} {\bibinfo {author} {\bibfnamefont {R.}~\bibnamefont
  {Gherghescu}}, \bibinfo {author} {\bibfnamefont {J.}~\bibnamefont {Skalski}},
  \bibinfo {author} {\bibfnamefont {Z.}~\bibnamefont {Patyk}}, \ and\ \bibinfo
  {author} {\bibfnamefont {A.}~\bibnamefont {Sobiczewski}},\ }\href@noop {}
  {\bibfield  {journal} {\bibinfo  {journal} {Nucl. Phys. A}\ }\textbf
  {\bibinfo {volume} {651}},\ \bibinfo {pages} {237} (\bibinfo {year}
  {1999})}\BibitemShut {NoStop}%
\bibitem [{\citenamefont {Warda}\ and\ \citenamefont {Egido}(2012)}]{War12}%
  \BibitemOpen
  \bibfield  {author} {\bibinfo {author} {\bibfnamefont {M.}~\bibnamefont
  {Warda}}\ and\ \bibinfo {author} {\bibfnamefont {J.~L.}\ \bibnamefont
  {Egido}},\ }\href@noop {} {\bibfield  {journal} {\bibinfo  {journal} {Phys.
  Rev. C}\ }\textbf {\bibinfo {volume} {86}},\ \bibinfo {pages} {014322}
  (\bibinfo {year} {2012})}\BibitemShut {NoStop}%
\bibitem [{\citenamefont {Delaroche}\ \emph {et~al.}(2006)\citenamefont
  {Delaroche}, \citenamefont {Girod}, \citenamefont {Goutte},\ and\
  \citenamefont {Libert}}]{Del06}%
  \BibitemOpen
  \bibfield  {author} {\bibinfo {author} {\bibfnamefont {J.-P.}\ \bibnamefont
  {Delaroche}}, \bibinfo {author} {\bibfnamefont {M.}~\bibnamefont {Girod}},
  \bibinfo {author} {\bibfnamefont {H.}~\bibnamefont {Goutte}}, \ and\ \bibinfo
  {author} {\bibfnamefont {J.}~\bibnamefont {Libert}},\ }\href@noop {}
  {\bibfield  {journal} {\bibinfo  {journal} {Nucl. Phys. A}\ }\textbf
  {\bibinfo {volume} {771}},\ \bibinfo {pages} {103} (\bibinfo {year}
  {2006})}\BibitemShut {NoStop}%
\bibitem [{\citenamefont {Nazarewicz}(1993)}]{Nazarewicz1993489}%
  \BibitemOpen
  \bibfield  {author} {\bibinfo {author} {\bibfnamefont {W.}~\bibnamefont
  {Nazarewicz}},\ }\href {\doibase
  http://dx.doi.org/10.1016/0375-9474(93)90565-F} {\bibfield  {journal}
  {\bibinfo  {journal} {Nucl. Phys. A}\ }\textbf {\bibinfo {volume} {557}},\
  \bibinfo {pages} {489 } (\bibinfo {year} {1993})}\BibitemShut {NoStop}%
\end{thebibliography}%

\end{document}